\documentclass[prb,showpacs]{revtex4}
\usepackage{graphicx}
\usepackage{dcolumn}
\usepackage{bm}

\begin{document}

\title{Electron transport of a quantum wire containing a
finite-size impurity under THz electromagnetic field illumination}

\author{Guanghui Zhou$^{1,2,3}$}
\email{ghzhou@hunnu.edu.cn}
\author{Yuan Li$^2$}
\author{Fang Cheng$^2$}
\author{Wenfu Liao$^2$}

\affiliation{$^1$CCAST (World Laboratory), PO Box 8730, Beijing
             100080, China}
\affiliation{$^2$Department of Physics, Hunan
             Normal University, Changsha 410081, China\footnote{Mailing address}}
\affiliation{$^3$International Center for Materials Physics,
             Chinese Academy of Sciences, Shenyang 110015, China}

\begin{abstract}
We theoretically investigate the electron transport properties for
a semiconductor quantum wire containing a single finite-size
attractive impurity under an external terahertz electromagnetic
field illumination in the ballistic limit. Within the effective
mass free-electron approximation, the scattering matrix for the
system has been formulated by means of a time-dependent mode
matching method. Some interesting properties of the electron
transmission for the system have been shown through a few groups
of numerical examples. It is found that in the case of the
comparative stronger field amplitude and the frequency resonant
with the two lowest lateral energy levels in the impurity region,
the field-induced intersubband transition dominates the process as
if without the impurity. And there is a step-arising on the
transmission as a function of the incident electron energy.
However, in the case of lower field amplitude and the non-resonant
frequencies both multiple symmetry Breit-type resonance peaks and
asymmetry Fano-type dip lines appear in the electron transmission
dependence on the incident energy due to the presence of the
impurity and the external field. Therefore, within certain energy
range the transmission as a function of the field frequency and/or
field amplitude shows a rich structure. Moreover, the transmission
dependence on the strength and size of the impurity is also
discussed. It is suggested that these results mostly arise from
the interplay effects between the impurity in a quantum wire and
the applied field.

\end{abstract}
\pacs{73.23.-b; 73.21.Hb; 78.67.Lt}

\maketitle

\section{Introduction}
Quantum devices using a two-dimensional electron gas (2DEG) formed
in high-mobility GaAs/Al$_x$Ga$_{1-x}$As semiconductor
heterostructure have been found to have better performance
characteristics than bulk devices. It has been predicated that the
characteristics of the devices may be further improved by using
quantum wires in which the carriers behave as a
quasi-one-dimensional gas. And some technological success has been
achieved in recent years in growing high-quality semiconductor
quantum wires with a length up to $\mu$m.$^1$ In the ballistic
regime and at low temperatures quantum coherent effects will
dominate the electron transport properties of a mesoscopic system.
One of the most important features is that, when the lateral size
of a quantum wire varies, the conductance shows an histogram
structure and each step has an height of $2e^2/h$ or integer times
of it for both short wire$^2$ (quantum point contact) and long
wire.$^{1,3}$

The electron transport properties of the quantum wire formed on a
2DEG can be affected by many factors. The presence of disorders in
a quantum wire generally leads to a suppression of the conductance
plateaus below integer values.$^4$ Especially, for a quantum wire
containing an attractive impurity the transmission shows resonance
dips below each confinement subband.$^{5-7}$ Also, the interaction
of electrons induces transport anomalies.$^8$ However, there has
been growing interest in the time-dependent transport for quantum
wire systems in recent years, such as presence of a time-modulated
potential$^9$ and quantum pumping.$^{10}$ Further, when a quantum
wire is illuminated under an external electromagnetic (EM) field,
due to the inelastic scattering of electrons by photons many new
features have been observed experimentally$^{11,12}$ and predicted
theoretically.$^{13-15}$ The technique of applying an external
field is of particular interest, since no additional current and
voltage probes have to be attached to the sample which may disturb
the system's properties. More recently, a new type of giant
magnetoresistance has been discovered$^{16}$ and theoretically
explained$^{17}$ in a high-mobility 2DEG subject to a high
frequency microwave radiation and a vertical magnetic field. It is
therefore of great interest in basic physics aspect to study the
time-dependent transport properties of quantum structures on
semiconductor 2DEG system. On the other hand, possible
applications of nanostructures in future electronic devices, which
will have to operate at very high frequencies, require detailed
knowledge of their frequency- and time-dependent transport
behavior.

The values of the lateral energy level separation and the Fermi
energy are of the order of 1$\sim$100meV for the typical
semiconductor quantum wires. This corresponds to frequencies of
the range of 0.25$\sim$25 terahertz (THz), which is available in
experiments with the development of the ultrafast laser technology
and the physics in this frequency range is of interest now.$^{18}$
When the Fermi level is below the lowest lateral level at the
bottleneck part of a quantum wire, electrons can not go through
without the assistance of an external EM field. However, under the
field illumination, electrons in the wire can absorb energy of
photons and go through this geometric barrier (the neck).
Therefore, in the region of barrier the electron transmission is
determined by the combined effect of the external EM field and the
wire lateral shape variation.$^{11-14}$ Recently, with an emphasis
on the pure EM field effect on the electron transport, we have
theoretically investigated$^{19}$ a straight (uniform
cross-section) and clean (without impurity) quantum wire
illuminated under a transversely polarized THz EM field. An
interesting transmission step-like structure has been predicated
when the EM field frequency is resonant with the two lowest
lateral energy levels of the quantum wire. Also, the impurity
effect on the transport is important for a thin and long quantum
wire because the electron interaction effect is observable in the
presence of backscattering.$^{1,3}$ Without considering this
interaction, the resonance structures of the electron transmission
in a straight quantum wire with a finite-size scatter for a wide
range of the impurity parameters have been investigated in the
absence of external EM field.$^7$ In the present paper, we combine
the models in Refs.[7,19] and study the electron transmission
through a straight quantum wire containing a finite-size
attractive impurity under a transversely polarized THz EM field
illumination. In this case the transmission behavior may reflect
the interplay effects between the impurity$^7$ and the external EM
field$^{19}$ on the electron transport. Within the effective mass
free-electron approximation, the scattering matrix for the system
has been formulated through a time-dependent mode matching
method.$^{19}$ Using a group of numerical examples we demonstrate
some interesting electron transmission behaviors for this system.
To the best of our knowledge, the interplay effect between the
impurity and the external EM field on the electron transport for a
quantum wire has not been reported previously. This effect may be
important for the understanding of basic physics in
low-dimensional systems and for the future nanoscale circuit
applications.

The outline of the paper is as follows. In Sec. II we set up the
problem for a straight quantum wire containing a finite-size
impurity under an external THz EM field illumination in terms of a
single-electron time-dependent Schr\"{o}dinger equation, and
calculate the electron transmission probability through the system
by the time-dependent mode matching in the framework of
Landauer-B\"{u}ttiker formalism. Some numerical examples to
illustrate the dependence of the electron transmission on the
incident energy, field parameters, and impurity parameters
respectively are presented and discussed in Sec. III. Finally,
Sec. IV gives a conclusion of the paper.

\section{Model and Formalism}
The system under study is an ideal straight 2D quantum wire
(quantum waveguide) of width $D$, containing a finite-size
attractive impurity, which is depicted in Fig. 1
schematically.$^{7,19}$ The quantum wire smoothly connects the two
electron reservoirs (leads) at each end. The x-axis is
longitudinally along the wire, and the y-axis describes the
transverse direction. The impurity range $0\le x\leq l$ in the
quantum wire is illuminated under a transversely polarized THz EM
field in an unspecified way. The field vector potential can be
described as ${\bf A}=(\varepsilon/\omega)\cos(\omega t)\hat e_y$
with angular frequency $\omega$ and amplitude $\varepsilon$ ($\hat
e_y$ is the unit vector in the polarized direction).

Within an effective mass approximation, the single-particle
time-dependent Schr\"{o}dinger equation in the field illuminated
impurity region is
\begin{equation}
i\frac{\partial}{\partial
t}\Psi(x,y,t)=\left[-\frac{\partial^2}{\partial
x^2}+(-i\frac{\partial}{\partial
y}+eA)^2+v_c(y)+v_i(x,y)\right]\Psi(x,y,t),
\end{equation}
where we have adopted the unit of $\hbar=2m^*=1$. In the
Hamiltonian, $v_c(y)$ presents a transverse confining potential in
the form of either a hard-wall or parabolic one which confines
electrons to the wire and to the reservoirs, and
\begin{equation}
v_i(x,y)=\theta(x)\theta(l-x)v_i(y)=-v\theta(x)\theta(l-x)\theta(d/2-\mid
y-y_c\mid)
\end{equation}
is used to represent the single finite-size impurity potential,
where $v$ is the strength of the impurity and $\theta(x)$ is the
step function (see Fig.1). This type of the finite-size single
impurity has been used to model an unintentional Be doping in GaAs
semiconductor$^5$ and may be artificially created in the quantum
waveguide using recent nanotechnology.$^6$

Using the method used in Ref.[19] (also see Ref.[20]) to solve the
time-dependent Schr\"{o}inger equation (1), and then considering
the scattering of the two interfaces between the impurity region
(with field illumination) and the clean region (without field
illumination) separately.$^{21}$ The detail knowledge about this
aspects is referred to Refs.[19,20] and is not presented here.

When an electron with total incident energy $E$ emits from the
left reservoir to the left interface at $x=0$, transmission and
reflection will take place simultaneously. Because the electron
has certain probability of absorbing a photon after penetrate the
interface, transition from the lower mode to the upper mode
happens. So there are two energy components of E and
$E+\hbar\omega$ in the reflected wave in the region of $x<0$
\begin{eqnarray}
\Psi(x,y,t)=[e^{i(k_1x-Et)}+c_1e^{-i(k_1x+Et)}]\phi_1(y)+
c_2e^{-i[k_2x+(E+\omega)t]}\phi_2(y),
\end{eqnarray}
where $\phi_n(y)$ (n=1,2) are transverse eigenfunctions with
eigenvalues $\epsilon_n$ in the clean region (without impurity),
$c_1$, $c_2$ are the reflection coefficients of the two modes
respectively, and
\begin{equation}
k_1=\sqrt{E-\epsilon_1},\;\;\; k_2=\sqrt{E-\epsilon_2 +\omega}
\end{equation}
are their associated wavevectors. Consequently, we can obtain the
electronic wavefunction in the region of $x>0$
\begin{eqnarray}
\Psi(x,y,t)=[c_+e^{i(k_+x-Et)}+c_-e^{i(k_-x-Et)}]\psi_1(y)+
[G_+c_+e^{i[k_+x-(E+\omega)t]}+G_-c_-e^{i[k_-x-(E+\omega)t]}]\psi_2(y),
\end{eqnarray}
where $\psi_n(y)$ and $\epsilon_n'$  are the solutions for the
transverse equation with impurity potential $v_i(y)$. $c_+$, $c_-$
are the transmission coefficients of the two field-spilt modes and
the constants $G_{\pm}=\pm(\sqrt{\gamma^2+\xi^2}\mp\gamma)/\xi$
(where $\gamma=\omega-(\epsilon_2'-\epsilon_1')$ is the detuning,
and $\xi$ is the two-mode coupling constant). The two electron
wavevectors in the impurity region are
\begin{eqnarray}
k_{\pm}=\sqrt{E-\epsilon_1'+\gamma/2\mp\sqrt{\gamma^2+\xi^2}/2}.
\end{eqnarray}

We can match the above two wavefunctions (3) and (5) at the
interface of $x=0$. Continuously connecting the two above
wavefunctions and their differentials gives the following four
algebraical equations for the coefficients $c_1$, $c_2$, $c_+$ and
$c_-$
\begin{eqnarray}
\lambda_{11}(1+c_1)=c_++c_-,\nonumber
\end{eqnarray}
\begin{eqnarray}
\lambda_{22}c_2=c_+G_++c_-G_-,\nonumber
\end{eqnarray}
\begin{eqnarray}
\lambda_{11}(k_1-c_1k_2)=c_+k_++c_-k_-,\nonumber
\end{eqnarray}
\begin{equation}
-\lambda_{22}c_2k_2=c_+G_+k_++c_-G_-k_-,
\end{equation}
where
$\lambda_{nn'}=\int_{-D/2}^{D/2}dy\psi_n^*(y)\phi_{n'}(y)\;(n,n'=1,2)$
is the matrix elements for connecting the two sets of transverse
eigenfunctions. With the solution of these algebraic equations in
Eq. (7) both the transmission and the reflection matrix for the
left interface (as if without the right interface) can be
expressed by
\begin{equation}
t'=\left[\matrix{\sqrt{k_+/k_1}\;c_+ & 0 \cr \sqrt{k_-/k_1}\;c_-&
0}\right],\;\;\;\; r=\left[\matrix{c_1 & 0 \cr \sqrt{k_2/k_1}\;c_2
& 0}\right].
\end{equation}

When we consider the electron transmission probability through the
whole real impurity, we use the approach recently developed in
Ref. [21] for the symmetric system. This approach needs to derive
the total scattering matrix which can be expressed in the
transmission matrix and reflection matrix on each interface. The
total transmission matrix is just the anti-diagonal submatrix of
the total scattering matrix in the symmetry system case.

Because of the similarity of the two interfaces one has not to
match the wave functions at the right interface $x=l$, but we need
to know the transmission and reflection matrix of electron
emitting from right to left for the left interface. In this case
the electronic wavefunction in the impurity region is
\begin{eqnarray}
\Psi(x,y,t)=[c_{+}^{e}e^{-ik_{+}x}+c_{+}^{e}e^{-ik_{-}x}+c_{+}^{r}e^{ik_{+}x}
+c_{-}^{r}e^{ik_{-}x}]e^{-iEt}\psi_{1}(y)\nonumber\\
+[G_+c_{+}^{e}e^{-ik_{+}x}+G_-c_{+}^{e}e^{-ik_{-}x
}+G_+c_{+}^{r}e^{ik_{+}x}+G_-c_{-}^{r}e^{ik_{-}x}]e^{-i(E+\omega
)t}\psi_{2}(y),
\end{eqnarray}
where $c_{\pm}^{e}$ are the coefficients of the electron emitting
from right to left, $c_{\pm}^{r}$ are the associated reflection
coefficients. Correspondingly, the transmitted electron
wavefunction in the region of $x<0$ is
\begin{eqnarray}
\Psi(x,y,t)=c_{1}^t e^{-i(k_{1}x+Et)}\phi_{1}(y)+c_{2}^t
e^{-i[k_{2}x+(E+\omega)t]}\phi_{2}(y),
\end{eqnarray}
where $c_{1}^t$ and $c_{2}^t$ are the transmission coefficients.
These two wavefunctions also satisfy the continuous condition at
$x=0$, from which we can obtain the transmission and reflection
matrix from right to left for this interface, $r'$ and $t$,
respectively (Here we do not present the detailed expressions for
them because they are much more complicated than Eq. (8) for $t'$
and $r$). Consequently, the total transmission matrix through the
two interfaces (the whole system) is
$t_{tot}=S_{12}=t(1-Xr'Xr')^{-1}Xt'$. Therefore, according to
Landauer-B\"{u}ttiker's formulation$^{22}$ the total electron
transmission probability through the whole system is
\begin{equation}
T=Tr[t_{\displaystyle tot}^\dag t_{\displaystyle tot}],
\end{equation}
where $X$ is the transfer matrix between the two interfaces of the
system.$^{19}$

\section{Results and Discussion}
In this section, we numerically calculate the transmission
probability from Eq. (12). The physical quantities of the system
are chosen to be a high mobility GaAs/Al$_x$Ga$_{1-x}$As
heterostructure$^{1,2}$ with a typical electron density
$n=2.5\times10^{11}$cm$^{-2}$ and $m^*=0.067~m_e$ (where $m_e$ is
the free electron mass). We choose the hard-wall transverse
confining potential and the width of the wire $D=500$\AA~ (see
Fig. 1) such that the unit of energy
$E^*=\epsilon_1=\hbar^2\pi^2/(2m^* D^2)=14.1$meV which corresponds
to the unit of time $t^*=\hbar/E^*=4.7\times10^{-14}$s.
Correspondingly, the field frequency unit $\omega^*=1/t^*=21.3$THz
and the amplitude unit $\varepsilon^*=22.1$V/cm. This kind of
radiation is available in experiments now.$^{18}$ We also use the
length unit $l^*=D/\pi=63.7$\AA.

For the hard-wall confining potential with a finite-size impurity
of $d=0.1D$, $l=25$, $y_c=0.17D$ (see Fig. 1) and strength
$v=6.3\epsilon_1$, the transverse eigenvalues in the impurity
region are solved by the quantum perturbation method as
\begin{equation}
\epsilon_1'=-0.23\epsilon_1,\;\;\;\epsilon_2'=3.34\epsilon_1
\end{equation}
with associated eigenfunctions $
\psi_1(y)=0.951\phi_1(y)-0.314\phi_2(y)$ and
$\psi_2(y)=0.314\phi_1(y)+0.951\phi_2(y)$. And different
transverse eigenvalues as well as eigenfunctions can be obtained
by variation of the impurity size.

In the following we systematically present some numerical
examples, while restrict our attention to the energy range
($\epsilon_1,\epsilon_2$) throughout the work.

\subsection{The characteristics of transmission dependence on incident energy}
First of all, in the absence of the EM field, the transmission
probability $T$ as a function of incident energy $E$ should be the
same as that in Ref. [7], which shows the Berit-Winger resonances
in the lower energies $\epsilon_1<E<\epsilon_2'$ and the multiple
asymmetric Fano lines in the upper energies
$\epsilon_2'<E<\epsilon_2$.

In the presence of the EM field, we show the calculated $T$ versus
$E$ for two combinations of the field parameters in Fig. 2 with
impurity size of $d=0.1D$, $l=25$, $y_c=0.17D$ and strength
$v=6.37$. The dashed line presents the case of resonant field
frequency $\omega=3.57(\gamma=0, \omega=\epsilon_2'-\epsilon_1'$)
with $\varepsilon=9.96$, while the solid line gives a
representative case of nonresonant frequency
$\omega=2(\gamma\neq0$) with $\varepsilon=2.97$. In the case of
the field frequency resonant (matching) with the energy spacing of
the two lowest levels, the transmission probability is similar to
that for the case of absent impurity$^{19}$ except for a
suppression of average 0.1. Also a step-raising of transmission
occurs at energy $E=\epsilon_1'+\xi/2=2.52$ (for these field
parameters the mode coupling constant $\xi=5.5$).$^{23}$ This
interesting phenomenon can be similarly explained by the
field-induced intersubband transition.$^{19}$ When an electron
penetrate though the interface, the transverse levels of the
electron in the field illuminated region are dressed and one
electron mode is split into the two time-dependent modes with the
longitudinal momentum $k_+$ and $k_-$, respectively. When
$E<\epsilon_1'+\xi/2$, $k_+$ is imaginary and its corresponding
mode is an evanescent (nonpropagating) mode which contributes
nothing to the transmission so that the total transmission
probability is suppressed to an half value. Further, when
$E>\epsilon_1'+\xi/2$ the both modes become propagating and all
contribute to the transmission. However, with the non-resonant
field frequency the structure of transmission probability is much
different from the case without impurity. For this case of
non-resonant frequency and weaker field amplitude we note that an
interesting asymmetry Fano-type resonance dip appear at
$E\approx2$, which indicates the formation of a quasibound
states.$^7$ Electrons with this particular incident energy can
make a transition from a propagating state to a quasibound state
by emitting an energy of $\hbar\omega$. So this resonance dip may
be a result from the feature of the impurity-induced quasibound
state. We can also explain this phenomenon with the mode
propagational property. Around the point of $E\sim 2$, from Eqs.
(4) and (6) we find that the wavevector $k_2\sim0$ and the
field-split electron wavevector $k_+\sim0$. This indicates that
the two modes $k_2$ and $k_+$ may be nonpropagating. So that $T$
reaches a nonzero minimum because $k_1$ and $k_-$ are always
propagating.

We also note that there are some resonance oscillations on the two
transmission curves in Fig.2. These oscillations physically result
from the interference of the forward- and backward-going electron
waves induced by the two interfaces of the impurity along the
transport direction. The resonance peaks with prefect transmission
on the solid line of Fig. 2 appeared at lower energies
$E\approx$1.01, 1.39 and 1.85 may be identified to the symmetry
Breit-type resonance and the half width of each resonance
specifies the lifetime of the corresponding quasibound state.
Therefore, the above characteristics of $T$ dependence on $E$ for
the system have implied the differences in electron transmission
between the two cases of with and without$^7$ EM field and also
between the two cases of with and without$^{19}$ impurity.

\subsection{Transmission dependence on the field parameters}
Next, in this subsection we investigate the influence of the field
parameters on the electron transmission with the interesting
energy of $E=2$ for the nonresonat frequency case (see Fig. 2). We
present the numerically calculated T as a function of $\omega$ and
$\varepsilon$ as shown in Fig. 3 with the same impurity parameters
as that in Fig. 2. From Fig. 3 we can see that the transmission
probability $T$ changes between $0$ (black) and $1$ (white) with
the variation of the field parameters and shows several dip
structures (black areas) around $\omega\sim2$, which includes the
dip of the solid line in Fig. 2 at $\varepsilon\sim3$. We also
note that there exist perfect transmission areas (complete white).
In the region of small $\varepsilon$ (regardless of the value of
$\omega$) the field is too weak to affect the transmission, while
in the region of $\omega>3.57$ (resonant) regardless of the value
of $\varepsilon$ the field-induced intersubbund transition
dominates. However, with the proper combinations of $\varepsilon$
and $\omega$ perfect transmission also occurs. This interesting
phenomenon of perfect and blocked transmission has also been
predicted$^{14}$ for a narrow-wide-narrow shape clean quantum wire
under a EM field illumination for certain combinations of the
field parameters.

In the following, we discuss the effects of $\omega$ and
$\varepsilon$ on $T$ separately. Fig.4 shows the calculated $T$ as
function of $\omega$ for four different values of $\varepsilon$
with the same impurity parameters as that in Fig.2. In the case of
a rather weak field amplitude $\varepsilon=1.5$ as shown in Fig.
4(a), multiple Fano resonance lines appear within
$\omega=1.69\sim1.82$. This result is very similar to that in
Ref.[7] but within the range of higher energies. From Eqs. (4) and
(6), we know that in this case mode $k_+$ begins to open
(propagating) earlier than $k_2$ with $\omega\approx1.69$ and
$\varepsilon=1.5$ ($\xi=1.75$). There are two propagating modes
($k_+$ and $k_-$) in the impurity region but only $k_1$ is
propagational outside the impurity. The multiple Fano resonances
are connected with the interaction of multiple quasidonor levels
but lowered by the EM field. In Fig. 4(b) with $\varepsilon=2.97$
(same value as that in solid line of Fig. 2) there is only one
Fano dip at $\omega=2$, which corresponds to the single dip of the
solid line in Fig. 2. Furthermore, as shown in Fig. 4(c) with
stronger amplitude $\varepsilon=6$, $T$ is suppressed with no dip
in the range of $2.2<\omega<3$ because the open of $k_2$ enhances
the probability of reflection. And $T$ increases when $\omega>3$
due to the second mode $k_+$ beginning to be propagating. So it is
reasonable when only one mode ($k_-$) in the illuminated region
with $\varepsilon=10$ is propagating and $T$ is more suppressed as
shown in Fig. 4(d).

On the other hand, in Fig. 5 we present the calculated $T$ as
function of $\varepsilon$ for four different values of $\omega$
with the same impurity parameters. For a rather small $\omega=1$
in Fig. 5(a) $T$ shows an oscillation behavior with almost
periodic maximum (perfect transmission) and declined minimum
(valley) as $\varepsilon$ increases. The oscillation results from
the interference of the forward- and backward-going electron waves
induced by the two interfaces of the impurity along the transport
direction. With a fixed $\omega=2$, the multiple Fano resonance
lines appear as shown in Fig. 5(b) within
$\varepsilon=0.8\sim2.6$. The pattern of $T$ is slightly different
from that of Fig. 4(a) but with the same physical reason. The
perfect transmissions are seen to occur at $\varepsilon$=0.78,
1.4, 1.68 and 2.3 and the asymmetric Fano dips are seen to occur
at $\varepsilon$=0.8, 1.38, 1.82 $\cdots$. The second mode ($k_+$)
is the evanescent mode and the coupling of the two modes in field
region vanishes as well as the formation of the quasibound states
when $\varepsilon>2.97$ holds. From the above analysis it seems
reasonable that the step structure appears around
$\varepsilon=3.4$ in Fig. 5(c) with $\omega=2.3$. This results
from the suppression of the second mode ($k_+$) with increasing of
$\varepsilon$ when $\omega>2$ holds. In Fig. 5(d) with $\omega=4$
the average $T$ is much enhanced with no dip or step structures
because both $k_+$ and $k_2$ are propagating modes.

In general for a fixed electron incident energy, a different
combination of the field parameters results in a different
transmission dependence. But for the chosen geometrical parameters
in this work the rich structure of the transmission always appear
in the ranges of $E\sim2$, $\omega\sim1.5-2$ and
$\varepsilon\sim1-3$.

\subsection{Transmission dependence on the impurity parameters}
Finally, in this subsection we investigate the effects of the
impurity parameters on the transmission probability $T$ for the
same system.

We first consider the effect of impurity size on the transmission.
An example of different impurity width $d=0.2D$ has been adopted
and its corresponding transverse eigenvalues have been calculated
as $\epsilon_1'=-1.85$ and $\epsilon_2'=3.19$. Fig. 6 shows the
numerical results of $T$ as a function of $E$ with a fixed
$\omega=2$ for two different $\varepsilon$. With
$\varepsilon=2.97$ the multiple Fano resonances appear in Fig.
6(a) within a energy range of $E=1.6\sim1.8$. However, Fig. 6(b)
shows that as $\varepsilon$ increases to 3.6 with the same
$\omega$, a single Fano dip line (similar to the solid line in
Fig. 2) appears. This indicates that the two modes ($k_2$ and
$k_+$) begin to be propagating. Comparing with the solid line in
Fig. 2 we conclude that as the width of the impurity increases, it
need stronger field amplitude $\varepsilon$ to suppress the
splitting of the second mode. But this conclusion may not be
extended to the increase of length $l$.

Next, we consider the effect of the impurity strength on the
transmission. An example of different impurity strength of
$v=8.0\epsilon_1$ has been adopted and its associated eigenvalues
have been recalculated as $\epsilon_1'=-0.64$ and
$\epsilon_2'=3.26$. Fig. 7 shows the numerical results of $T$ as a
function of $E$ with a fixed $\omega=2$ for two different
$\varepsilon$. A asymmetry Fano dip also occurs in Fig. 7(a) at
$E\approx1.9$ (with $\varepsilon=2.97$) and in Fig. 7(b) at
$E\approx2$ (with $\varepsilon=3.2$). From this result we conclude
that with the same $\omega$ although the increase of the impurity
strength one need stronger field amplitude $\varepsilon$ to
suppress the splitting of the second mode. It seems that the
variation of the impurity strength does not change the
transmission behavior much even if we use a $\delta$ potential to
model the impurity. And this fact of model independence is
physically reasonable.

\section{Conclusion}
We have theoretically investigated the electron transport
properties for a straight semiconductor quantum wire containing a
single finite-size attractive impurity under a THz EM field
illumination in the ballistic limit. Within the effective
free-electron approximation, a single-particle time-dependent
Schr\"{o}dinger equation was established and the scattering matrix
for the system was formulated via the method of time-dependent
mode matching.

The numerical examples predicate that a step-arising on the
transmission probability versus the electron incident energy
occurs in the case of the field frequency resonant with the
lateral energy spacing of the two lowest levels. This situation is
similar to the case of the system without$^{19}$ impurity and the
physical origin is mainly the coherent field-induced intersubband
transition. However, due to the interplay between the
impurity-induced quasibound states and the applied field, both
multiple symmetry Breit-type resonance peaks and asymmetry
Fan-type dip lines appear within the energy range of
($\epsilon_1$,$\epsilon_2$) in the case of a weaker field
amplitude and the non-resonant field frequency. This situation is
also consistent with the system$^7$ without field illumination.
Further, the dependence of the transmission on either field
amplitude or frequency also indicates the change of the shape and
the position of the resonance dips. But the rich structure of the
transmission always appears with the proper combination of the
field parameters. Moreover, the transmission behavior is almost
independent on the impurity parameters.

Therefore, from the results of this work we conclude that the
field parameters $\omega$ and $\varepsilon$ can control the
characteristics of electron transmission through the propagational
property of the modes in a quantum wire. These effects of the
applied field on the transport properties in a quantum wire may be
useful for understanding basic physics of quantum structures and
for device physics.

\begin{acknowledgements}
This work was supported by the Nature Science Foundation of Hunan
(NO. 02JJY2008) and by the Research Foundation of Hunan Eduction
Commission (NO. 04A031).
\end{acknowledgements}

\newpage
\begin{figure}
\center
\includegraphics[width=3in]{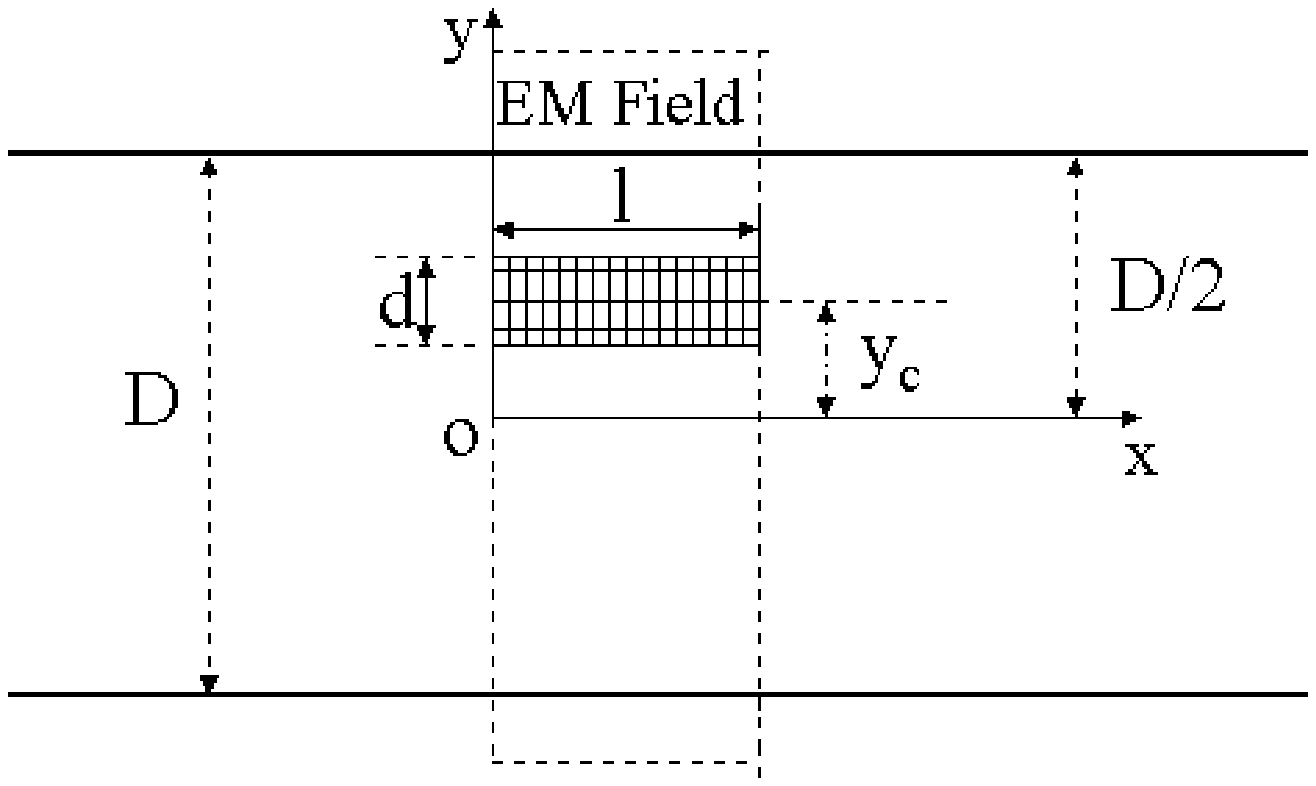}
\caption{Sketch of the system}
\end{figure}

\begin{figure}
\center
\includegraphics[width=3in]{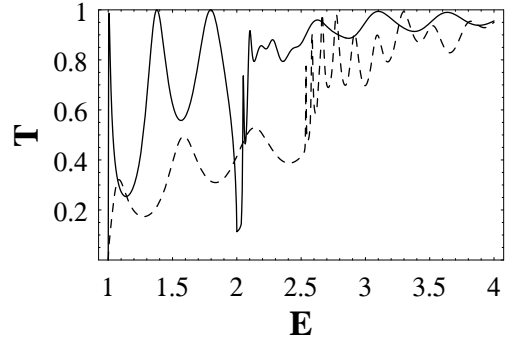}
\caption{Transmission probability T dependence on the incident
energy E (in units of $\epsilon_1$) for two combinations of the
field parameters in the energy range ($\epsilon_1,\epsilon_2$),
the dashed line for $\omega=3.57$ ($\gamma=0$, resonant case) and
$\varepsilon=9.96$, while the solid line for $\omega=2$ and
$\varepsilon=2.97$. We have used the parameters of $D=200\AA$,
$d=0.1D$, $l=25$ and $y_c=0.17D$ such that $\epsilon_1'=-0.23$ and
$\epsilon_2'=3.34$.}
\end{figure}

\begin{figure}
\center
\includegraphics[width=3in]{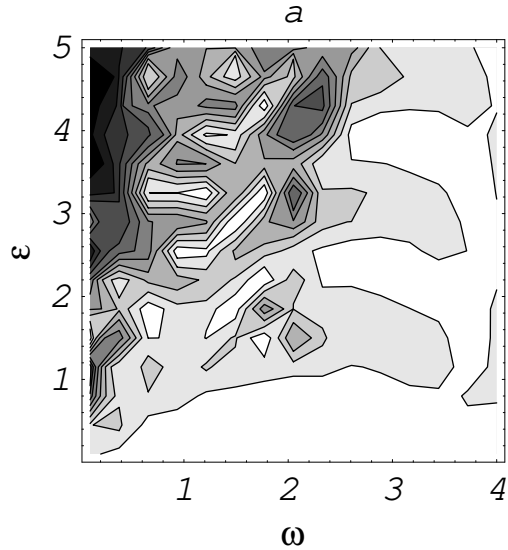}
\caption{Transmission probability T dependence on both field
frequency $\omega$ and amplitude $\varepsilon$ with incident
energy $E=2$, the system parameters are the same as in Fig. 2.}
\end{figure}

\begin{figure}
\center
\includegraphics[width=2.5in]{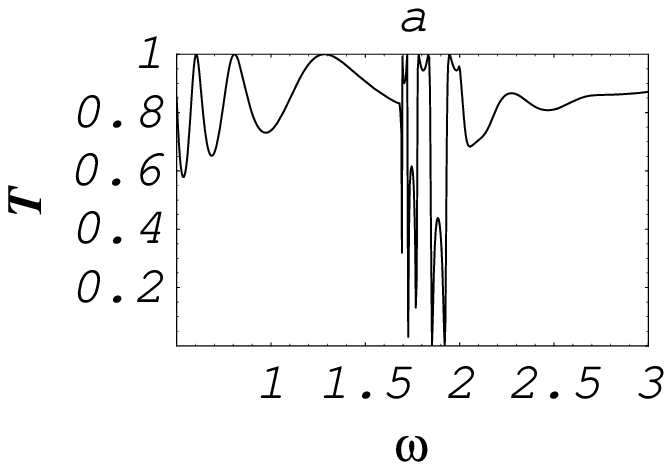}
\includegraphics[width=2.5in]{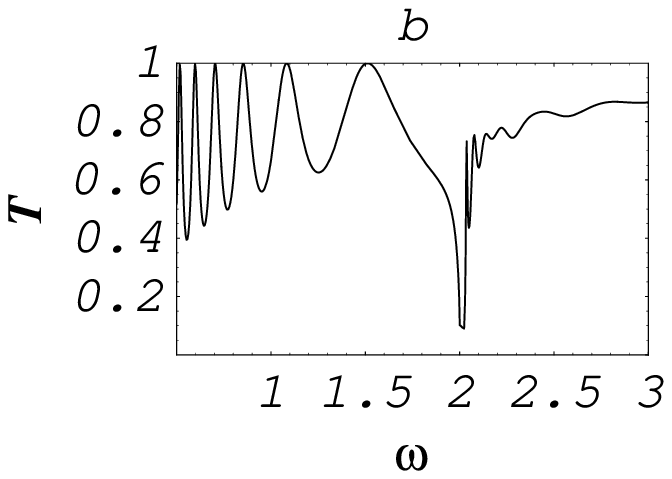}
\includegraphics[width=2.5in]{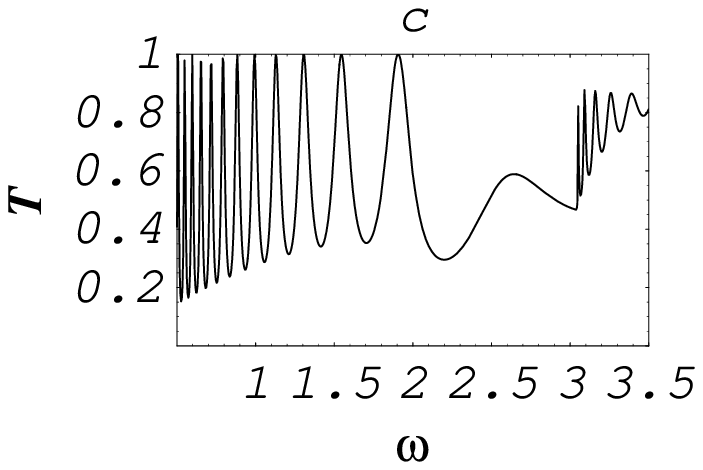}
\includegraphics[width=2.5in]{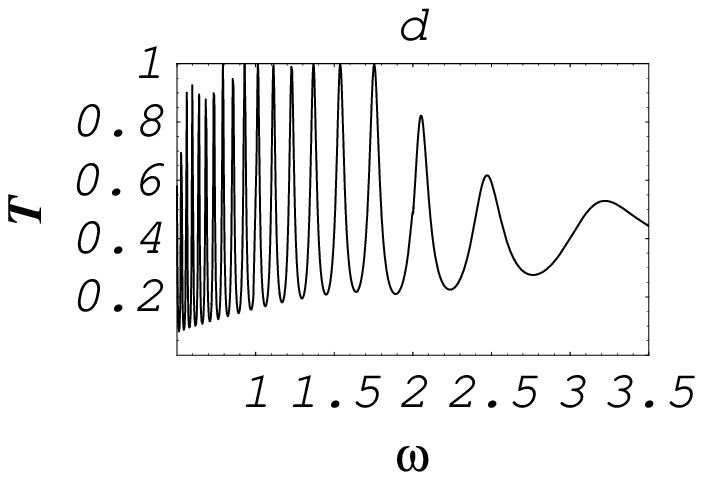}
\caption{Transmission probability T dependence on field frequency
$\omega$ with incident energy $E=2$ for serval field amplitudes:
(a) $\varepsilon=1.5$, (b) $\varepsilon=2.97$, (c) $\varepsilon=6$
and (d) $\varepsilon=10$. The system parameters are the same as in
Fig. 2.}
\end{figure}

\begin{figure}
\center
\includegraphics[width=2.5in]{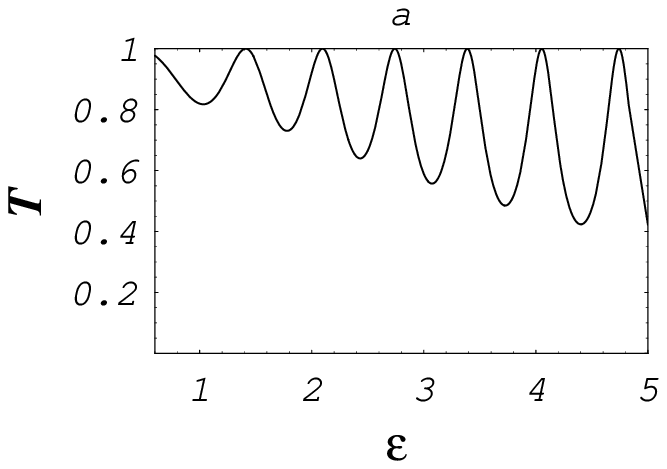}
\includegraphics[width=2.5in]{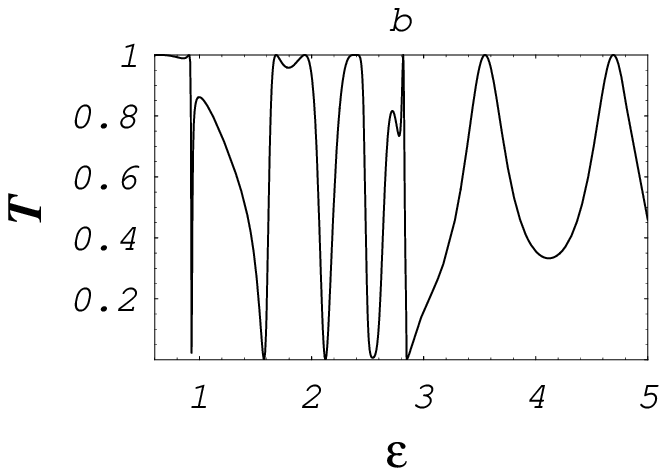}
\includegraphics[width=2.5in]{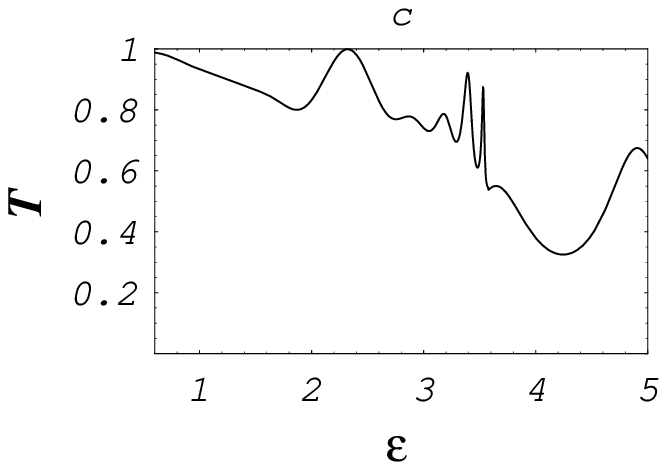}
\includegraphics[width=2.5in]{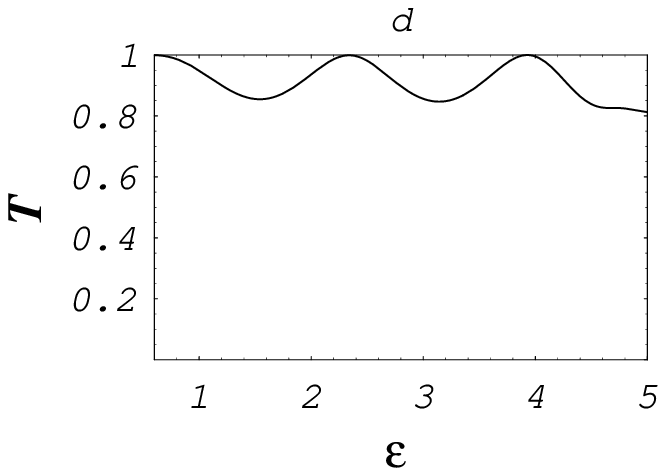}
\caption{Transmission probability T dependence on field amplitude
$\varepsilon$ with incident energy $E=2$ for serval frequency: (a)
$\omega=1$, (b) $\omega=2$, (c) $\omega=2.3$ and (d) $\omega=4$.
The system parameters are the same as in Fig. 2.}
\end{figure}

\begin{figure}
\center
\includegraphics[width=2.5in]{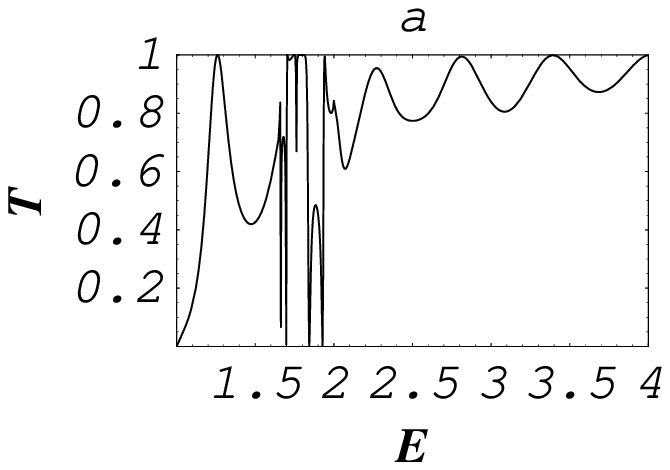}
\includegraphics[width=2.5in]{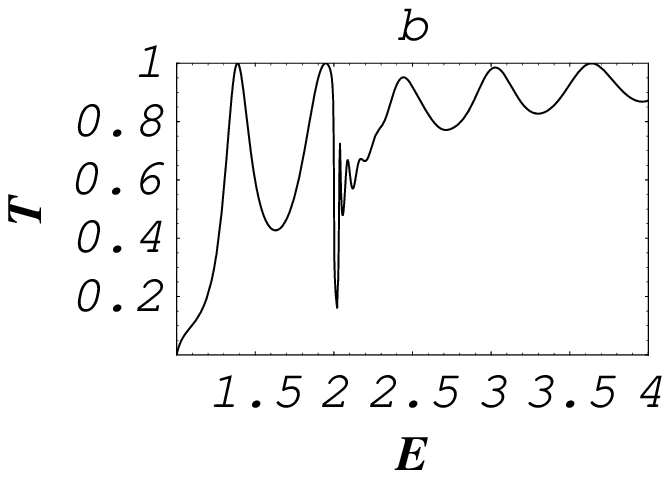}
\caption{Transmission probability T dependence on incident energy
E for a different impurity width of $d=0.2D$ with field frequency
$\omega=2$: (a) $\varepsilon=2.97$ and (b) $\varepsilon=3.6$. With
the same length $l$ and strength $v$ as in Fig. 2 the change of
the impurity width induces $\epsilon_1'=-1.85$ and
$\epsilon_2'=3.19$ }
\end{figure}

\begin{figure}
\center
\includegraphics[width=2.5in]{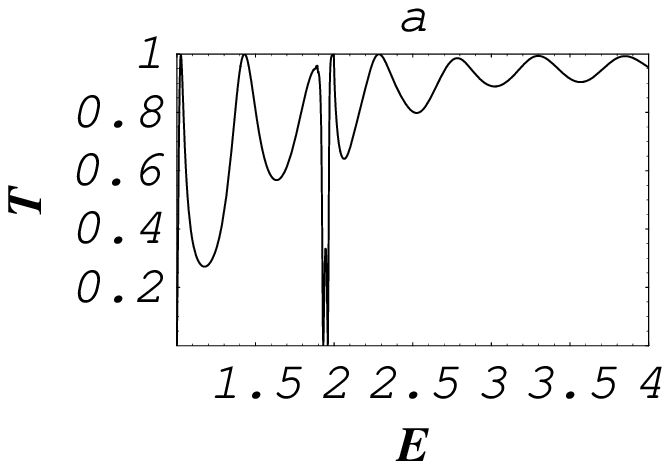}
\includegraphics[width=2.5in]{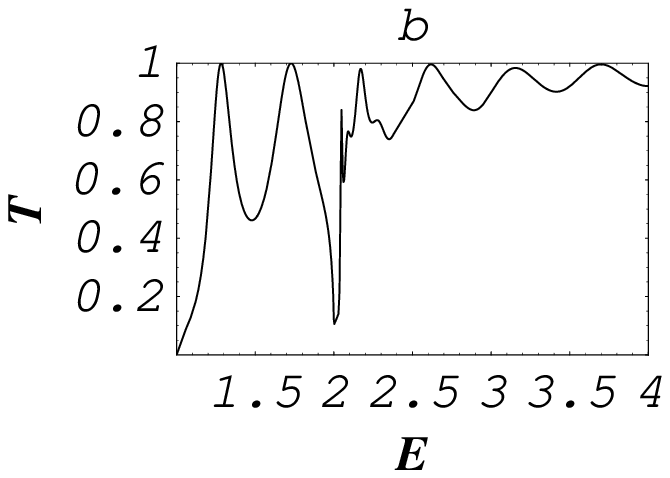}
\caption{Transmission probability T dependence on incident energy
E for a different impurity strength of $v=-8$ with field frequency
$\omega=2$: (a) $\varepsilon=2.97$ and (b) $\varepsilon=3.2$. In
this case $\epsilon_1'=-0.64$ and $\epsilon_2'=3.26$ with the same
impurity size as in Fig. 2.}
\end{figure}


\begin{references}
\bibitem{23}In the resonant case of $\gamma=0$, with the system parameters
selected in this work $\xi=5.5$. According to Eq. (6), $k_+>0$
requires that $E>\epsilon_1'+\xi/2=2.52$.
\end{references}
\end{document}